\documentclass[5p]{elsarticle}

 \usepackage{graphicx}
 \usepackage[font=footnotesize]{subfig} 

\usepackage{amssymb}
\usepackage{color}

\def\beq{\begin{equation}}
\def\eeq{\end{equation}}
\def\bea{\begin{eqnarray}}
\def\eea{\end{eqnarray}}

\newcommand*{\eqref}[1]{Eq.~(\ref{eq:#1})}
\newcommand*{\eqlab}[1]{\label{eq:#1}}
\newcommand*{\figref}[1]{Fig.~\ref{fig:#1}}
\newcommand*{\figlab}[1]{\label{fig:#1}}

\def\VYP#1#2#3{{\bf #1}, #3 (#2)}  

\begin{document}

\begin{frontmatter}



\title{On the feasibility of RADAR detection of high-energy neutrino-induced showers in ice}

\author[1]{Krijn D. de Vries}
\ead{krijndevries@gmail.com}
\author[2]{Kael Hanson}
\author[2]{Thomas Meures}
\address[1]{Vrije Universiteit Brussel, Dienst ELEM, B-1050 Brussels, Belgium}
\address[2]{Universit\'e Libre de Bruxelles, Department of Physics, B-1050 Brussels, Belgium}

\begin{abstract}
In this article we try to answer the question whether the radar detection technique can be used for the detection of high-energy-neutrino induced particle cascades in ice. A high-energy neutrino interacting in ice will induce a particle cascade, also referred to as a particle shower, moving at approximately the speed of light. Passing through, the cascade will ionize the medium, leaving behind a plasma tube. The different properties of the plasma-tube, such as its lifetime, size and the charge-density will be used to obtain an estimate if it is possible to detect this tube by means of the radar detection technique. Next to the ionization electrons a second plasma due to mobile protons induced by the particle cascade is discussed. An energy threshold for the cascade inducing particle of 4~PeV for the electron plasma, and 20~PeV for the proton plasma is obtained. This allows the radar detection technique, if successful, to cover the energy-gap between several PeV and a few EeV in the currently operating neutrino detectors, where on the low side IceCube runs out of events, and on the high side the Askaryan radio detectors begin to have large effective volumes.
\end{abstract}

\end{frontmatter}

\section{Introduction}
Recently the IceCube neutrino observatory~\cite{I3_2013sc} for the first time showed the existence of high-energy cosmic neutrinos. Since neutrinos are not deflected by magnetic fields in our universe, they should point back to their original source. This opens a new field in physics, the field of neutrino astronomy. At the highest energies, these cosmic neutrinos are extremely rare. At energies above several PeV, IceCube runs out of events and an even larger detector volume than the 1~km$^3$ covered by IceCube is needed for their detection. Due to the long attenuation length of the radio signal, the radio detection technique is an excellent candidate to detect these rare events.

Several radio detectors have been developed to detect the radio emission from neutrino-induced particle cascades in ice and moon rock~[2-10]. These are based on the emission from a net electron excess which develops when the particle cascade evolves, the Askaryan effect~\cite{Ask62,Zas92,Mun97}. The Askaryan radio-emission mechanism has been confirmed experimentally at SLAC~\cite{Sal01} and in the radio emission from air showers~\cite{Sch12,Mar11}. The Askaryan radio detection experiments have been developed to detect the GZK neutrino flux~\cite{Grei66,Zat66}, which should arise from the interaction of high-energy protons ($E>10^{19.5}$~eV) interacting with the Cosmic Microwave Background. Therefore, these detectors start to have large effective volumes for cascade inducing particles having energies in the EeV region and above, where the GZK flux is expected. It follows that there is an energy gap between IceCube, which is sensitive below several PeV, and the Askaryan radio detectors which start to have large effective volumes at EeV energies. In this article, we discuss the radar detection technique as a possible method to bridge this important energy region between several PeV and a few EeV.


The concept of radar detection of cosmic-ray-induced particle cascades in air dates back to the 1940s of the previous century. Blacket and Lovel~\cite{Bla40} proposed to use the radar detection technique to measure these cosmic-ray-induced air showers. Initial experimental attempts using the radar technique were done, but no conclusive evidence for the detection of air showers was found. It would take another 50 years before the interest in this subject was renewed~\cite{Bar93,Gor01}. This triggered several new modeling attempts~\cite{Bak10,Tak11,Sta13} and experiments
[25-30]. Even-though a first possible detection of a cosmic-ray-induced air shower might have been observed~\cite{Tak10}, no conclusive evidence for such a detection has been obtained so-far. Next to the efforts done for the radar detection of cosmic-ray air showers, recently suggestions were made to measure the reflection of radio waves from particle cascades induced in rock salt and ice~\cite{Chi13}.

With the existing infrastructure already available at the different Askaryan radio detection sites such as ARA~\cite{ARA} and ARIANNA~\cite{ARIANNA}, in this article, we discuss the radar detection technique for the detection of high-energy cosmic neutrinos. An energy threshold for the primary cascade inducing particle is derived for coherent scattering of the over-dense plasma region. The over-dense plasma region  is defined by the condition that the detection frequency is below the plasma frequency, where the plasma frequency scales with the electron density. In this regime, the incoming radio signal does not penetrate the plasma and scatters of the surface of the plasma tube. This brings a great advantage of ice as a medium over air. The volume in which the particle cascade is confined decreases dramatically in ice, resulting in higher plasma frequencies. It should be noted however, that it is also possible to scatter of the individual electrons in the under-dense plasma. Currently, most of the existing radar facilities for the detection of air showers are based on the detection of the under-dense plasma.

In the first section, we discuss the particle cascade and the induced ionization plasma. We discuss results obtained experimentally by irradiating ice with 3~MeV electrons and X-rays, where it is found that next to the ionization electrons, a long-lived plasma exists which is attributed to free protons~\cite{Ver78,Kun80,Haa83a,Haa83b}. In the following we use the experimentally obtained lifetime of these plasmas to determine an energy threshold for the radar detection of the over-dense plasma region. Finally, we conclude by calculating the radar return power for the different components of the plasma. This allows us to determine the maximum detection range for different values of the radar power considering two different cascade geometries.

\section{The plasma}
When a high-energy cosmic neutrino interacts in the medium a cascade of secondary particles is induced. To model the electromagnetic cascade we use a Heitler model \cite{Hei54}, stating that every interaction length $\lambda$, the total number of particles doubles and their average energy is split. This goes on up to the critical energy where the brems-strahlung, and creation-annihilation cross-sections become small compared to the ionization cross-sections. The critical energy of electrons in ice and their radiation length is given by,
\bea
E_c&=&0.0786\;\mathrm{GeV}\\
X_0&=&36.08\;\mathrm{g/cm^2}.\\
L_0&=&\frac{36.08\;\mathrm{g/cm^2}}{0.92\;\mathrm{g/cm^3}}=39.22\;\mathrm{cm.}
\eea
Where the ice density is assumed to be constant and equal to $\rho_{ice}=0.92\;\mathrm{g/cm^3}$. Using the radiation length $X_0$, the interaction length is given by, $\lambda=X_0 \ln(2)=25.01\;\mathrm{g/cm^2}=27.19\;\mathrm{cm}$. Now following the Heitler model stating that every radiation length the total number of particles is doubled and their energy is split, we can make an estimate for the maximum number of particles in the shower and the shower length. The maximum number of particles in the cascade can be estimated by,
\bea
N_{max}&=&\frac{E_{p}[\mathrm{GeV}]}{E_c[\mathrm{GeV}]}\nonumber\\
&=& 12.72\;E_{p}[\mathrm{GeV}].
\eqlab{Heit}
\eea
A more realistic shower development is given by the NKG parameterization, developed by Kamata and Nishimura \cite{Kam58}, and Greisen~\cite{Grei65}. It follows (see Appendix A) that the maximum number of particles in the Heitler model is overestimated by a factor of 10 independent of energy in the TeV to EeV energy range. Therefore, in the following we will consider,
\beq
N_{max}^{NKG}=1.27\;E_{p}[\mathrm{GeV}].
\eeq
The shower maximum is reached after $n=\ln(E_p/E_c)/\ln(2)$ divisions, leading to an estimate for the shower length of,
\bea
l_c&=&n\lambda\nonumber\\
&=&L_0\;\ln(E_p/E_c)\nonumber\\
&\approx&0.4\;\ln(12.72\;E_{p}[\mathrm{GeV}])\;\mathrm{m}.
\eqlab{length}
\eea
This is the length it takes for the cascade to reach its maximum, the total length will therefore in general be at least a factor of two longer. In~\figref{fig1}, the shower length is plotted as a function of primary energy from 1 TeV to 1 EeV and is found to be of the order of 5-10 meters. This corresponds well to the values obtained using the NKG parameterization (see Appendix A). It should be noted that we ignore the LPM effect~\cite{Lan35,Mig57} which becomes significant above PeV-EeV energies~\cite{Alv97,Alv98}, giving rise to an increased shower length.
\begin{figure}[!ht]
\centerline{
\includegraphics[width=.5\textwidth, keepaspectratio]{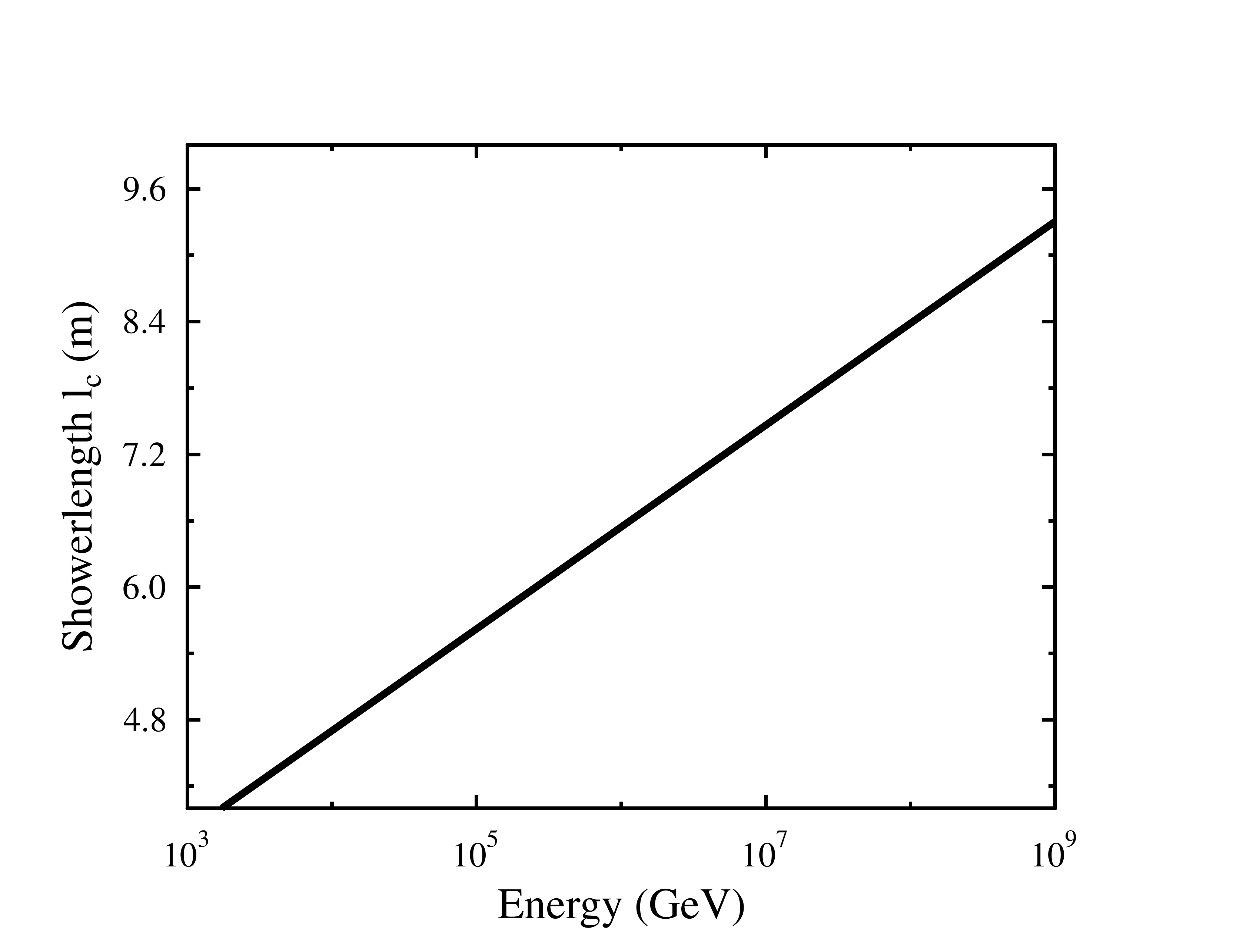}}
\caption{The shower length $l_c$ given in~\eqref{length}, as a function of the energy of the primary shower inducing particle. }
\figlab{fig1}
\end{figure}

So-far, we discussed the high-energy electron-positron pairs in the shower front. Nevertheless, we will not focus on this high energetic shower front for the radar detection of the cascade. This is mainly due to two reasons, one is that the particle density is too low, the second is that the shower front is very narrow such that we would need to measure at very high frequencies. 
More interesting are the low-energy ionization electrons left behind when the shower front has passed. The ionization energy of ice can be estimated as $E^{ionization}_{H_2O}\approx 20\;\mathrm{eV}$. Assuming that most of the energy loss of the electrons or positrons goes into ionization, the energy lost in one radiation length at the shower maximum equals $E^{lost}=E_{c}-(1/e) E_c\approx50$~MeV per high energy electron in the shower front. Hence the number of ionization electrons created by a single high energy electron equals $N^{ion}=E^{lost}/E^{ionization}_{H_2O}/L_0=6.93\cdot10^4\;\mathrm{cm^{-1}}$. The total number of ionization electrons created is thus given by,
\bea
N_e^{tot}&&=N_{max}\cdot N^{ion}\nonumber\\
&&=8.80\cdot10^4\;E_p[\mathrm{GeV}]\;\mathrm{cm^{-1}}
\eea
To convert the line density to a volume density, we need to consider the radial particle distribution in the cascade. This is modeled by modifying the parameterization used in~\cite{Wer12}, which was originally developed for air showers,
\beq
w(r)=\frac{\Gamma(4.5-s)}{\Gamma(s)\Gamma(4.5-2s)}\left(\frac{r}{r_0}\right)^{s-1}\left(\frac{r}{r_0}+1\right)^{s-4.5},
\eeq
The shower age parameter is taken to be its value at the shower maximum, $s=1$, and the distance parameter $r_0^{ice}=7$~cm is converted using its value in air. More details can be found in Appendix B. In~\figref{figA2}, we plot the radial particle distribution.
\begin{figure}[!ht]
\centerline{
\includegraphics[width=.5\textwidth, keepaspectratio]{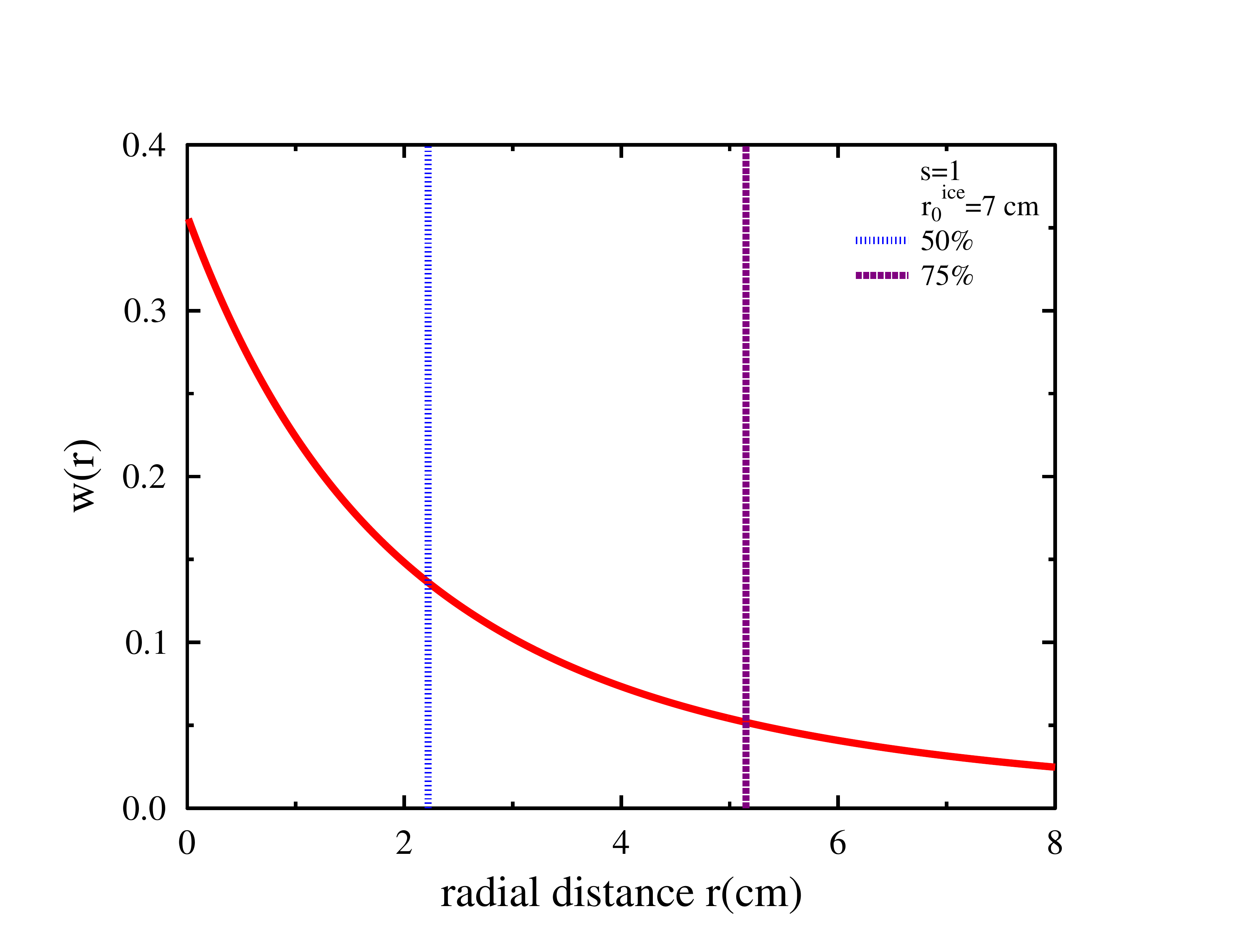}}
\caption{The radial particle distribution in the cascade following~\cite{Wer12}. The distance parameter $r_0^{ice}=7$~cm is converted from its approximate value in air equal to $r_0^{air}=80$~m. The shower age parameter $s=1$ is chosen equal to its value at the shower maximum. The blue dotted and purple dashed lines indicate the radius containing 50$\%$ and 75$\%$ of the particles. }
\figlab{figA2}
\end{figure}
It follows that the radial particle distribution is a fast decreasing function and hence also the particle density drops rapidly as function of radial distance. The blue dotted and purple dashed lines indicate the radius in which $50\%$ and $75\%$ of the particles are contained. In the following we only consider the inner 2~cm containing $\sim 50\%$ of the particles. Correcting for the total number of particles and the area of the ring, we obtain an ionization electron density of
\beq
n_e=N_e^{tot}\cdot 0.5/(\pi 2^2)\approx 3.5\cdot10^3\;E_{p}[\mathrm{GeV}]\;\mathrm{cm^{-3}}.
\eqlab{Ntot}
\eeq

\begin{figure}[!ht]
\centerline{
\includegraphics[width=.5\textwidth, keepaspectratio]{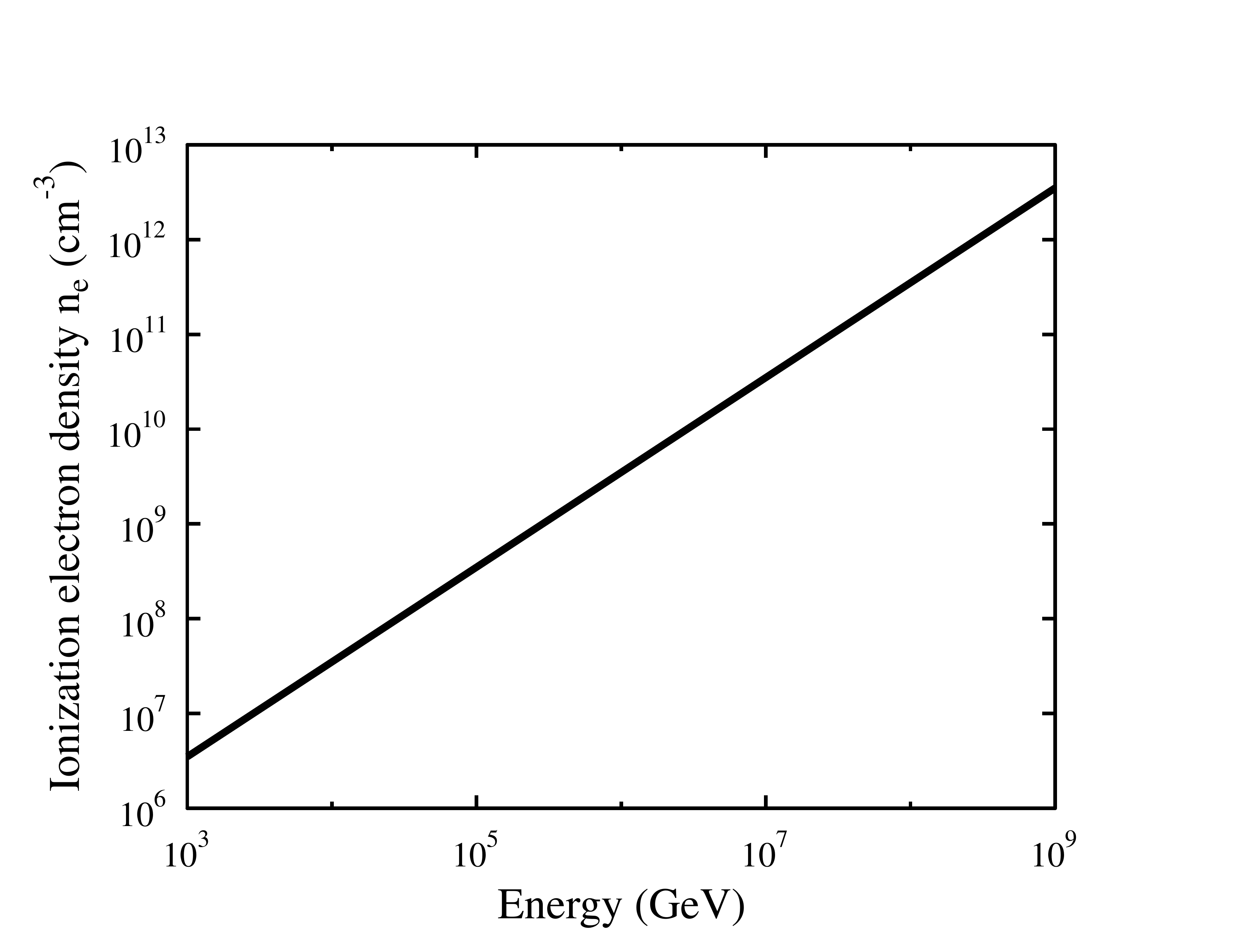}}
\caption{The ionization electron density as modeled by~\eqref{Ntot} as a function of the energy of the primary shower inducing particle. }
\figlab{fig2}
\end{figure}

\subsection{Experimental results}
The expected ionization electron density is given in \figref{fig2} and can be compared to the numbers obtained experimentally in~\cite{Ver78,Kun80,Haa83a,Haa83b}. These numbers are obtained by irradiating a block of ice with X-rays and 3 MeV electrons. The total number of free electrons was accordingly measured for an absorbed dose of 20~mrad in the $\sim\mathrm{cm^3}$ ice sample, corresponding to an integrated energy slightly larger than $10^{14}$~eV. The total number of conduction electrons was found to be $N_e^{tot}=(2\cdot10^{-11}\;\mathrm{M})\cdot (6.022\cdot 10^{23}\;\mathrm{N_e M^{-1}})\approx 10^{13}$ electrons. Since we consider a source moving at the speed of light, to compare the density, we have to use the dimensions of the plasma tube considered in the previous section. Furthermore, in the previous section we only considered the energy lost in a single radiation length for $50\%$ of the particles. Taking these factors into account, the corresponding density becomes $n_e\approx5\cdot 10^{9}\;\mathrm{cm^{-3}}$, which is slightly larger than the value obtained from~\figref{fig2}.

In~\cite{Ver78,Haa83a} it was also shown that the electron plasma has a lifetime of the order of 100~ps at relatively high temperatures close to the freezing point which increases to tens of nanoseconds at temperatures below -60$^\circ$ Celcius, also depending on the amount impurities in the ice. Even-though ice temperatures found for example at the South-Pole are rather low and of the order of -50$^\circ$ Celcius, in the following we will assume a conservative value of 1~ns. More interesting, next to the free (mobile) ionization electrons a conduction transient with a much longer lifetime of the order of 10~ns to $1\;\mathrm{\mu s}$ was observed~\cite{Kun80,Haa83b}. This conduction transient is attributed to free (mobile) protons, where the assumption was made that the total number of protons is equal to the total number of ionization electrons, $n_e=n_p$. In the following sections, we will keep this assumption.


\section{Radar detection of the electron plasma}
The most important properties to determine if radar can be used to detect a plasma are the plasma frequency, $\nu_{e}$, the plasma lifetime $\tau_{e}$, and the length of the plasma tube $l_c$. For all frequencies below the plasma frequency, the radio signal does not penetrate the plasma and scatters of the surface of the plasma tube. Next to the plasma frequency, also the lifetime and dimensions of the plasma cloud are important. If the lifetime of the plasma is smaller than $1/\nu_{e}$ or the length of the plasma tube is smaller than $c_{med}/\nu_e$, the speed of light in the medium divided by the plasma frequency, the plasma will not be able to make a full oscillation and re-emit the incoming signal. This gives us two conditions for the observation frequency $\nu_{obs}$, 
\bea
\nu_{obs} &<& \nu_e \\
\nu_e > \nu_{obs} &>&\left\{
 \begin{array}{l l}
 1/\tau_e\;\;\;\;\;\;\;\;\;(c_{med}\tau_e < l_c)\\
 c_{med}/l_c\;\;\;\;(c_{med}\tau_e > l_c)
 \end{array}
 \right.
\eea
where the plasma lifetime is given by the trapping time of the low energy ionization electrons, which was found to be of the order of $\tau_e=1$~ns. The speed of light in the medium is given by $c_{med}=c_{vac}/n\approx c_{vac}/1.7$ taking an index of refraction in ice equal to $n_{ice}=1.7$. 

In the case of the electron plasma, $c_{med}\tau_e < l_c$. Therefore, the second condition becomes, $\nu_e > \nu_{obs} > 1/\tau_e > 1\;\mathrm{GHz}$. The plasma frequency depends on the induced low-energy electron density in the medium determined in~\eqref{Ntot}, and is given by,
\bea
\nu_{e}&=&8980\sqrt{n_e}\;\mathrm{Hz}\nonumber\\
&\approx& 0.5\sqrt{E_p[\mathrm{GeV}]}\;\mathrm{MHz.}\nonumber\\
\eea
The energy threshold for radar detection of high-energy-neutrino induced showers in the medium can now easily be estimated. Putting the plasma frequency to its minimum value of 1~GHz, we obtain
\beq
E_p[\mathrm{GeV}]\gtrsim 4\cdot 10^6\;\mathrm{GeV}
\eeq
Putting a threshold of 4~PeV when measuring at 1~GHz. 

\section{Radar detection of the proton plasma}
Up to now we only considered the free ionization electrons in the plasma. The free electrons become trapped at lattice defects in the ice within the very short time span of 1~ns~\cite{Ver78,Haa83a}. After the decay of these free electrons, still a conductivity transient was observed with a lifetime of tens to hundreds of nanoseconds~\cite{Kun80,Haa83b}. This effect was attributed to free protons left behind in the medium. 

It follows that we can also expect a long-lived plasma with similar properties from a high-energy-neutrino induced shower in ice. In this situation the condition $c_{med}\tau_p < l_c$ is not fulfilled for the longest lifetimes and the plasma frequency is bounded by $\nu_p > \nu_{obs} > c_{med}/l_c$. Taking a conservative value of $l_c=5$~m, we obtain $\nu_p > \nu_{obs} > 36\;\mathrm{MHz}$. It follows that due to the longer lifetime of the plasma, it is possible to detect it at lower frequencies in the MHz band. What is the plasma frequency for these free protons? We can express the plasma frequency for a charge of mass $m_p$ with respect to the electron mass $m_e$ as,
\beq
\nu_{p}=\sqrt{\frac{m_e}{m_p}}\nu_e.
\eeq
For protons, the plasma frequency becomes,
\beq
\nu_{p}=1.2\cdot10^{-2}\sqrt{E_p[\mathrm{GeV}]}\;\mathrm{MHz}.
\eeq
Taking an observer frequency of $50$~MHz gives,
\bea
E_p[\mathrm{GeV}]&\gtrsim& 2 \cdot 10^7\;\mathrm{GeV}.
\eea
Leading to an energy threshold of approximately 20~PeV, which is slightly larger than for the free electrons.

\section{The radar return power}
To determine the feasibility for radar detection of high-energy neutrino induced particle cascades in the ice, we first need to determine the return power of a signal transmitted with power $P_t$. In the far-field, the radiated power measured at a distance $R$ from the transmitter will fall off isotropically like $(4\pi R^{2})^{-1}$. The maximum length of the plasma tube is given by the shower length $l_c$ shown in~\figref{fig1}. It follows that the maximum length which is probed is 10 meters, while the typical distance to the transmitter will be a few hundred meters to several kilometers fulfilling the far-field condition. We now consider a bi-static radar configuration with the transmitter located at the top of the ice-sheet. Using a beamed transmission into the ice in the direction of the receiving antennas, we gain a factor of 4 over isotropic transmission and the power drops like  $(\pi R^{2})^{-1}$. It should be noticed that due to the gradient of the ice density the signal will be bend into the ice preventing the direct signal from reaching the receiver, and we do not have to worry about separating the direct signal from a possible reflection. In case of reflection, the signal is absorbed over the effective area of the plasma $\sigma_{eff}$ and re-transmitted isotropically to the receiving antennas giving an additional factor $(4\pi R^{2})^{-1}$, after which the signal is detected over an effective area, $A_{eff}$. Furthermore, at high frequencies, we have to consider the attenuation of the signal in the medium. The received power in this situation is given by,
\beq
P_r=P_t\eta\frac{\sigma_{eff}}{\pi R^{2}}\frac{A_{eff}}{4\pi R^2}e^{-4R/L_\alpha},
\eeq
where following~\cite{Gor01} we insert an additional efficiency factor $\eta$ which is estimated to be $\eta=0.1$. The attenuation is determined by the frequency dependent attenuation parameter $L_{\alpha}(\nu)$. Note that we assumed the distance from the transmitting antenna to the plasma, $R$, equal to the distance from the plasma to the receiving antenna. The total distance covered by the signal is therefore given by $R_{tot}=2R$. 

This leads us with the determination of the effective area for isotropic transmission of the plasma tube, $\sigma_{eff}$, also called the radar cross-section, the effective area of the receiving antenna, $A_{eff}$, and the frequency dependent attenuation parameter $L_{\alpha}$. Following~\cite{Gor01,Cri65}, the maximum radar cross-section for a wave at normal incidence on a thin wire is estimated by,
\beq
\sigma_{eff}^{max}(\phi)\simeq\frac{\pi L^2 \cos^4\phi}{(\pi/2)^2+(\ln[\lambda/(1.78\pi r_c)])^2},
\eeq
which is a function of the polarization angle $\phi$, the radius of the particle distribution $r_c=2$~cm, and the length of the plasma tube $L=c\tau_e=0.3$~m for the electron plasma and $L=l_c\approx 5$~m for the proton plasma. At an angle $\theta$ away from normal incidence the radar cross-section is approximated by,
\beq
\sigma_{eff}(\theta,\phi)\simeq\frac{\lambda^2 \tan^2\theta \cos^4\phi}{16\pi\left[(\pi/2)^2+(\ln[\lambda/(1.78\pi r_c \sin\theta)])^2\right]},
\eeq
which is valid for angles between $60^\circ \leq \theta \leq 120^\circ$. 

In general the effective area of a single receiving antenna scales like $A_{eff}=\lambda^2/B$, where $B$ is a fixed constant. Therefore, we will put $A_{eff}=\lambda^2$ and absorb the constant $B$ into the efficiency $\eta=0.1$.   

This leaves us with the determination of the frequency dependent attenuation parameter $L_{\alpha}$ for radio waves in ice. In~\cite{Bar12} measurements of the electric field attenuation length for vertically transmitted radio signals at the Ross ice-shelf are described. This is done by measuring the attenuation over a round-trip distance of 1155~m. A parameterization of the attenuation parameter $L_\alpha$ is given by,
\beq
<L_\alpha>=469-0.205\nu+4.87\cdot10^{-5}\nu^2\;\;\;\mathrm{m},
\eqlab{att}
\eeq
where the frequency $\nu$ is given in MHz. The parameterization is based on measurements between 75 and 1250~MHz, therefore an extrapolation has to be made to determine $<L_\alpha>$ at 50~MHz. It should be noted that the obtained attenuation length is fitted specifically for the Ross ice-shelf, and differs for example from the attenuation length measured by the ARA collaboration at the South-Pole~\cite{ARA}, which is found to be around a kilometer at 300~MHz. 

\subsection{Skin effects}
One of the effects we did not consider so-far is the skin effect. An electromagnetic wave impinging a conducting material will be attenuated when penetrating the conductor. A measure for this attenuation is the skin depth, after which the flux density of the signal has dropped by a factor $1/e$. Neglecting energy dissipation due to collisions which typically occurs at a lower frequencies, the skin depth is given by~\cite{Kol97},
\beq
\delta=\frac{c}{2\nu_p},
\eeq
where $c$ denotes the speed of light and $\nu_p$ is the plasma frequency. A conductor becomes opaque when its size is significantly smaller than the skin depth. In opposite, a conductor becomes highly reflective when detecting at frequencies well below the plasma frequency, in the situation where the skin depth is small compared to the dimension of the plasma. 

Considering an isotropic particle density within the plasma tube, close to the determined energy threshold the skin depth will be large compared to the lateral dimensions of the plasma tube, decreasing rapidly toward higher energies. Therefore, additional loss in the reflected signal can be expected as function of the plasma frequency and hence of the energy of the primary cascade inducing particle. Nevertheless, this is under the assumptions of an isotropic particle distribution. In reality, this distribution is highly peaked toward small radial distances as can be seen in~\figref{figA2}, and it can be shown that at energies close to the threshold energy there will still be a highly reflective tube which is of smaller lateral dimensions than considered in our calculation. A detailed description of the efficiency loss due to skin effects goes beyond the scope of this paper and we intend to include this in a future work. Possible inefficiencies due to the skin effect will therefore be absorbed in our efficiency factor $\eta$, which thus in principle should depend on frequency. 

\subsection{Return power for the electron plasma}
We now have all ingredients to determine the return power for the electron plasma. We will consider two different situations, the ideal situation where the incoming wave is at normal incidence and the polarization angle is zero, and the non-ideal situation of a wave coming in at $\theta=60^\circ$ with a polarization angle of $\phi=60^\circ$. Furthermore, an observer frequency of 1~GHz as determined in Section 3 is used. The input parameters now become,
\bea
\lambda=c_{med}/(1\;\mathrm{GHz})&=&0.18\;\mathrm{m}\nonumber\\
\sigma_{eff}^{max}(\phi=0^\circ)&=&0.11\;\mathrm{m^2}\nonumber\\
\sigma_{eff}(\phi=60^\circ,\theta=60^\circ)&=&1.6\cdot10^{-4}\;\mathrm{m^2}\nonumber\\
L_\alpha&=&313\;\mathrm{m}.\nonumber\\
\eea
In~\figref{epow}, the return power $P_r$ is plotted for the two different situations as a function of the distance $R$ from the emitter to the plasma.
\begin{figure}[!ht]
\centerline{
\includegraphics[width=.5\textwidth, keepaspectratio]{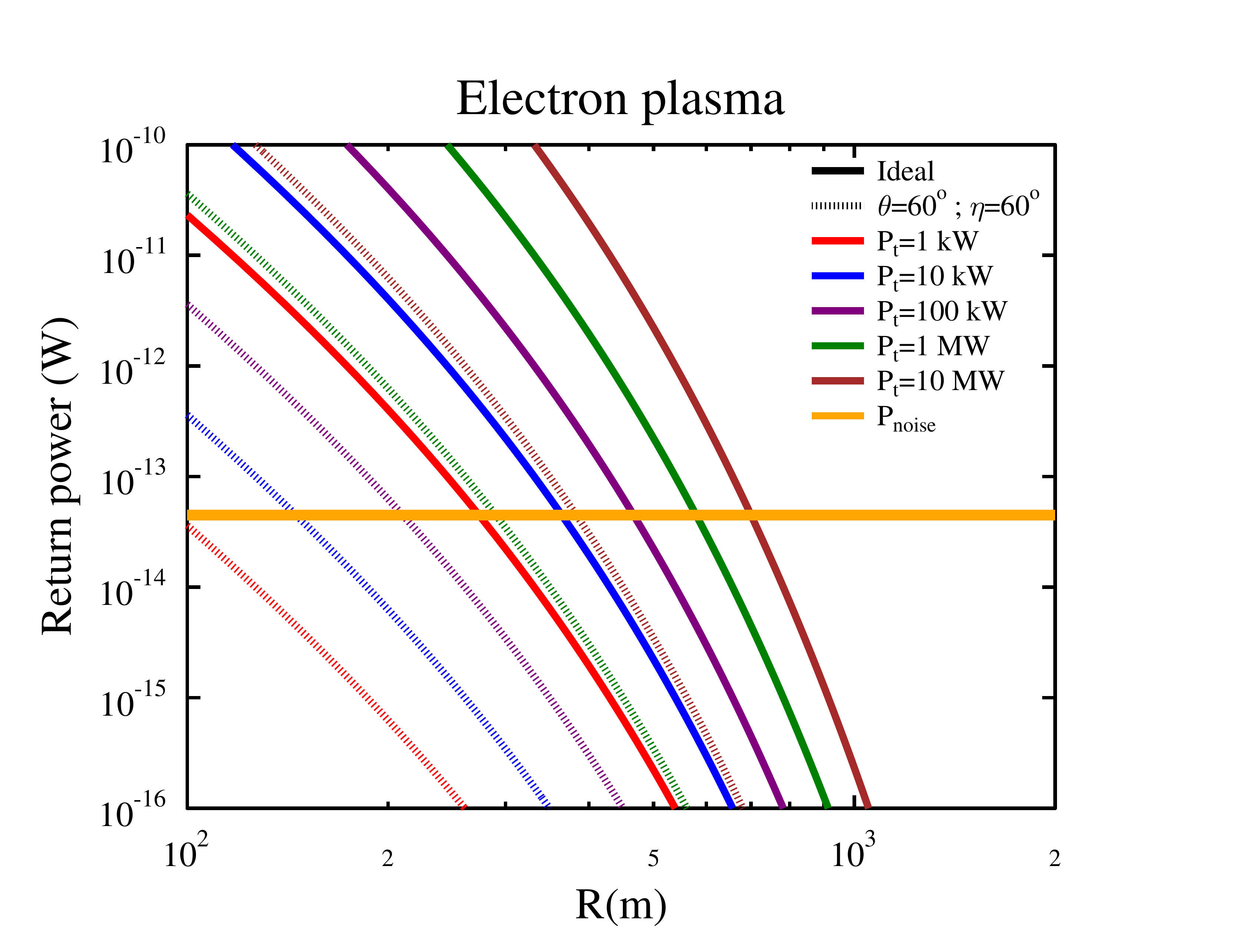}}
\caption{The return power as function of the source distance $R$ and transmitted power $P_t$. The power is given in the ideal situation of a signal at normal incidence and zero polarization angle, and the situation of a wave at $\theta=60^\circ$ incidence angle with the plasma and a polarization angle of $\phi=60^\circ$. The noise floor estimated by the thermal noise of the system is indicated by the horizontal orange line. }
\figlab{epow}
\end{figure}
The noise power is approximated by $P_{noise}=k_b\;T_{sys}\;\Delta \nu$, where $k_b$ is Boltzmann's constant, $T_{sys}=325$~K is chosen following in-ice measurements at the ARA site~\cite{ARA}, and the frequency band-width is approximated by $\Delta \nu=10$~MHz. 

It follows that the detection distance $R_{tot}=2R$, is limited to a few hundreds of meters in the non-ideal situation with a transmission power of $P_t=1$~kW up to approximately 1500~m in the ideal situation with a transmitted power of $P_t=10$~MW. 

The results are given for a single receiving antenna. For a grid of $N$ receiving antennas, the signal power scales coherently like $P_r^N=N^2\cdot P_r$, where the noise is incoherent and scales like $P_{noise}^{N}=N\cdot P_{noise}$ increasing the detection distance. 

\subsection{Return power for the proton plasma}
To determine the return power of the proton plasma, we again consider the two different situations described previously. The ideal situation where the incoming wave is at normal incidence to the plasma and the non-ideal situation where the angle of incidence as well as the polarization angle equals $60^\circ$. As determined in Section 4, the proton plasma can be detected at lower frequencies, leading to a longer attenuation length and an increased radar cross-section. The input parameters become,
\bea
\lambda=c_{med}/(50\;\mathrm{MHz})&=&3.6\;\mathrm{m}\nonumber\\
\sigma_{eff}^{max}(\phi=0^\circ)&=&5.5\;\mathrm{m^2}\nonumber\\
\sigma_{eff}(\phi=60^\circ,\theta=60^\circ)&=&1.2\cdot10^{-2}\;\mathrm{m^2}\nonumber\\
L_\alpha&=&459\;\mathrm{m}.\nonumber\\
\eea
\begin{figure}[!ht]
\centerline{
\includegraphics[width=.5\textwidth, keepaspectratio]{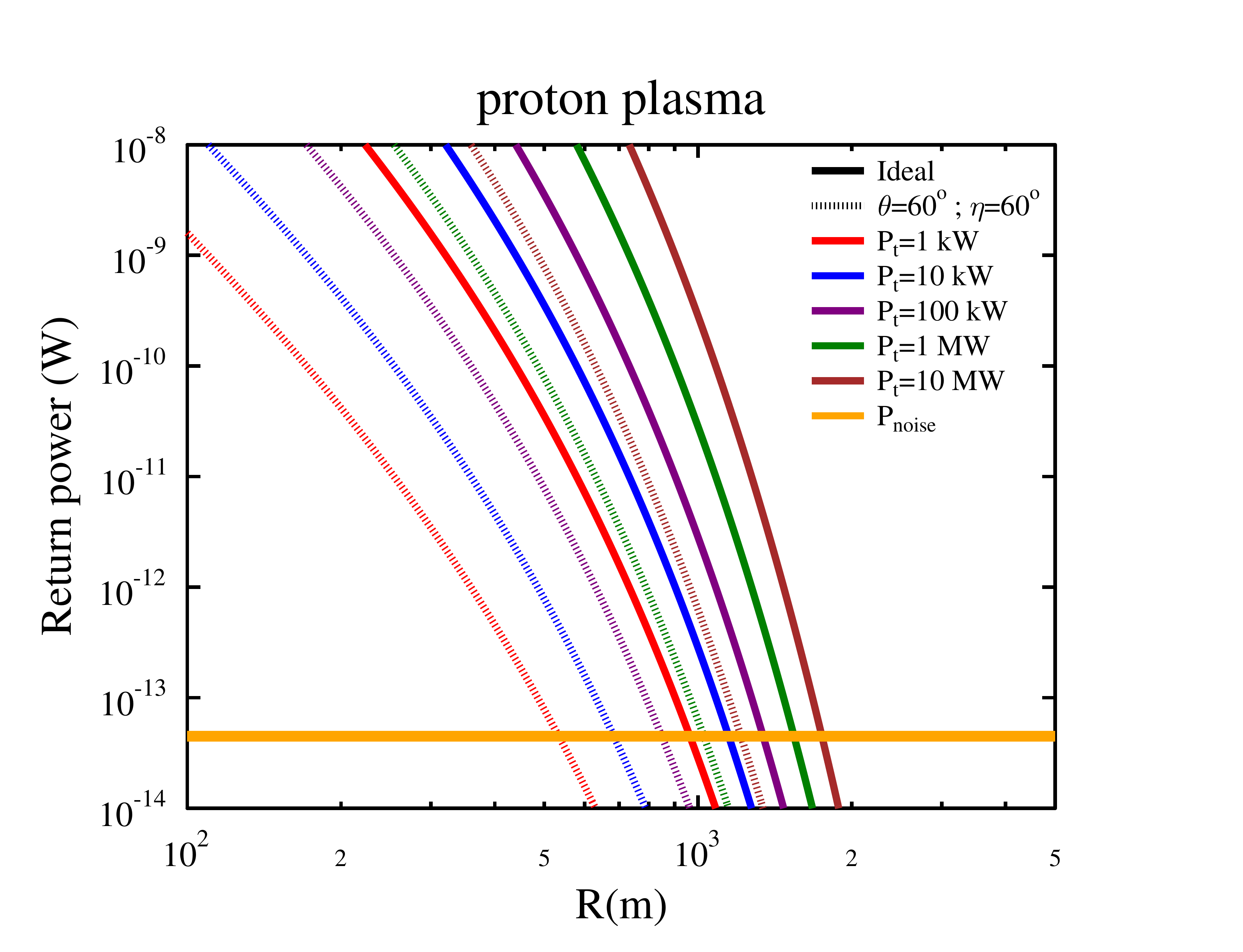}}
\caption{The return power as function of the source distance $R$ and transmitted power $P_t$. The power is given in the ideal situation of a signal at normal incidence and zero polarization angle, and the situation of a wave at $\theta=60^\circ$ incidence angle with the plasma and a polarization angle of $\phi=60^\circ$. The noise floor estimated by the thermal noise of the system is indicated by the horizontal orange line. }
\figlab{ppow}
\end{figure}
Using these parameters, in~\figref{ppow}, the return power $P_r$ is plotted as function of the source distance $R$ for different values of the transmitted power $P_t$. It follows that the detection distance $R_{tot}=2R$, is above 1~km even at the non-ideal situation for a low transmitted power of $P_t=1$~kW considering a single receiving antenna. 


\section{Discussion}

\subsection{The plasma}
The main uncertainty in the calculation are the different components of the ionization plasma in ice. We assumed a very short electron trapping time of 1~ns. Measured values obtained by~\cite{Ver78,Haa83a}, fluctuate between 100~ps at relatively high temperatures close to freezing point, to tens of nanoseconds at temperatures below -60$^\circ$ Celcius, also depending on the amount of impurities in the ice. Ice temperatures found for example at the South Pole, have relatively low values around -50$^\circ$ degrees Celcius, making our estimate of 1~ns conservative. In the case of the positive proton plasma, lifetimes measurements by~\cite{Kun80,Haa83b} were used. Following these measurements the number of protons was taken to be equal to the number of ionization electrons modeled. It follows that the obtained values strongly depend on the type of ice and the temperature. Repeating the measurements performed by~\cite{Ver78,Haa83a,Kun80,Haa83b} using ice obtained from the Antarctic ice sheet should provide more accurate values. 

The plasma frequency as modeled is given under the assumption that the electrons and protons are free and are not influenced by the medium. A measure for this is the mobility of the charges. In~\cite{Ver78,Haa83a}, the electron mobility was found to be $\mu_e\approx 25\;\mathrm{cm^2 V^{-1} s^{-1}}$ at ice temperatures between -60$^\circ$~C and -120$^\circ$~C. This is of the same order or even slightly larger than the electron mobility for a plasma in air. The measured proton mobility was about a factor $\sim10^3$ below the electron mobility which one would expect from the mass difference between the electrons and protons. Therefore, we can safely assume that the electrons as well as the protons are free and the used expression for the plasma frequency is valid.




\subsection{The radar return power}
For calculating the radar return power we determined several parameters. Furthermore, we fixed the geometry to a bi-static radar configuration. This led to an increase in the return power since we considered a beamed transmission over a quarter sphere. An additional factor may be gained by considering an even more beamed transmission. 

The attenuation length $L_{\alpha}$ was chosen conservatively following a parameterization from measurements performed at the Ross ice-shelf~\cite{Bar12}. Measurements performed at the ARA site~\cite{ARA} give attenuation lengths which are a factor of two larger than those used in our calculation.

Furthermore, we only considered a single receiving antenna, where for example the ARA~\cite{ARA} array currently operates 2 stations both composed out of 16 antennas consisting out of both horizontal and vertical polarization channels.  With respect to the background, the return power would gain a factor equal to the number of receiving antennas. In case of the ARA setup this would thus immediately give an additional factor of 30 for the return power with respect to the background noise.  


Finally, in calculating the radar return power, we ignored the under-dense scattering of the ionization plasmas, as well as the shower-front electrons. The component due to the under-dense scattering has been modeled for the detection of air showers~\cite{Bak10,Tak11,Sta13}. From these models it also follows that a strong boosting of the reflected signal of the shower-front electrons can be expected. Next to the boosting, depending on the geometry, there will be a frequency shift of the return signal which can be used to accurately determine the direction of the primary neutrino. A direct consequence of detecting the frequency shifted signal for different geometries is that it has to be detected over a rather wide frequency range, posing a technical challenge. A more complete model discussing different geometrical configurations, including the contribution of the under-dense regime in combination with effects like the frequency shift and boosting of the signal, is foreseen in a future publication.

\section{Conclusion}
The radar detection method as a probe for high-energy neutrino induced particle cascades in ice is discussed. Following the measurements described in~\cite{Ver78,Kun80,Haa83a,Haa83b}, two different components to the ionization plasma in ice are considered. The electron plasma, with a lifetime of approximately 1~ns, and a proton plasma with a longer lifetime between 10~ns and 1~$\mu s$. It follows that the plasma has to be probed with frequencies above 36~MHz for the proton plasma and frequencies larger than 1~GHz for the electron plasma. The plasma frequency for both components is derived as a function of the energy of the primary cascade inducing particle. This allows us to put an energy limit for the radar detection of the over-dense region by demanding that the plasma frequency has to be larger than the minimum observation frequency determined by either the lifetime or the size of the plasma. The energy limit for the electron plasma is estimated to be 4~PeV, where the energy limit of the proton plasma is estimated by 20~PeV.

Next to the energy threshold, the radar return power is calculated for a bi-static radar configuration. Comparing the return power to the thermal noise floor, we calculate the maximum detection distance for two different cascade geometries. The return power is given for different values of the transmitted radar power for a single receiving antenna. The detection distance for the electron plasma is estimated to be a few hundreds of meters in the non-ideal situation where the incoming wave has an incidence angle equal to the polarization angle of $60^\circ$, for a transmitted power of 1~kW. This increases up to several kilometer transmitting at 10~MW for the ideal situation where the incoming wave is normal to the plasma with zero polarization angle. The detection distance for the proton plasma is estimated to be larger than 1~km even for the non-ideal situation described above, and increases to several kilometers in the most ideal situation. 

It should be noted that the results are given for a single receiving antenna, and the detection distances will increase for a grid of antennas. Furthermore, conservative values were assumed for the radar cross-section and the attenuation length. This makes the radar detection technique a very promising method for the detection of high-energy neutrino induced particle cascades in ice in the up to now unexplored energy region between several PeV where IceCube runs out of events and a few EeV where the Askaryan radio detectors start to have large effective volumes.

\section{Acknowledgment}
The authors would like to thank O. Scholten for the very helpful discussions. We wish to thank the following funding agencies for their support of the research presented in this report: The Odysseus program of the Flemish Foundation for Scientific Research (FWO) under contract number G.0917.09., and the FRS-FNRS (Convention 4.4508.10 - Kael Hanson).

\begin{appendix}
\section{The cascade}
For the particle cascade, we considered a simplified Heitler model. The main advantage of this simplification is that we were able to derive the energy threshold for the different plasmas analytically. A more realistic parameterization is given by Kamata and Nishimura~\cite{Kam58} and Greisen~\cite{Grei65}. The shower is parameterized as,
\beq
N(X)=\frac{0.31\;\exp[(X/X_{0})(1-1.5\ln s)]}{\sqrt{\ln(E/E_{crit})}}
\eqlab{NKG}
\eeq 
where $X(g/cm^2)=0.01\cdot l\cdot\rho$ is the depth, given by integrating the density, $\rho$, along the path of the shower, $l(\mathrm{m})$. The density is modeled as a constant given by $\rho_{ice}=0.92\;\mathrm{g/cm^3}$. We use $E_c=0.0786$~GeV, and $X_0=36.08\;\mathrm{g/cm^2}$ for the critical energy and the electron radiation length in ice. The shower age $s$ is given by,
\beq
s(X)=\frac{3X/X_0}{(X/X_0)+2\ln(E/E_{crit})}.
\eqlab{age}
\eeq
\begin{figure}[!ht]
\centerline{
\includegraphics[width=.5\textwidth, keepaspectratio]{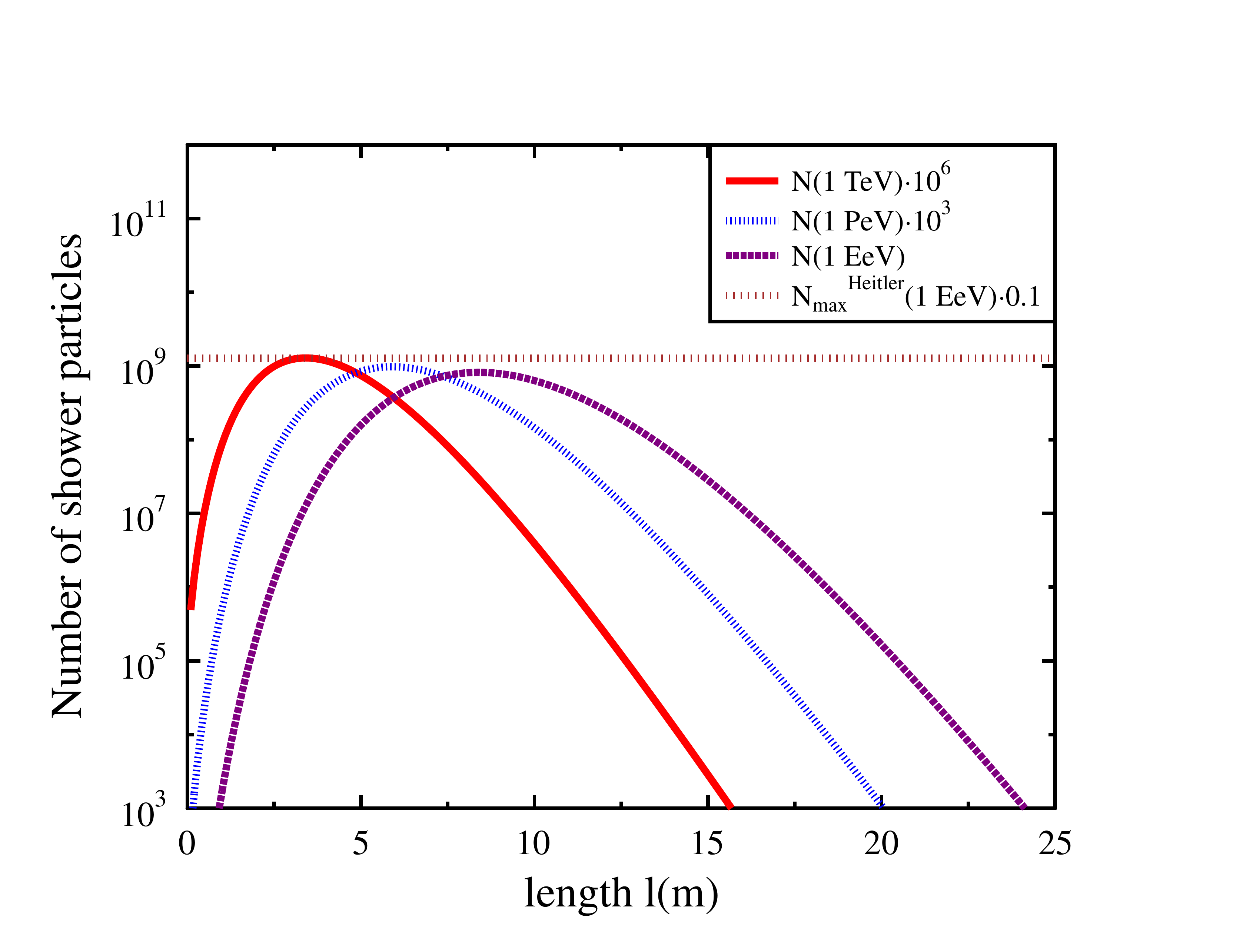}}
\caption{The total number of particles in the shower modeled using the NKG approximation~\eqref{NKG}. The cascade is modeled at three different energies, 1~TeV, 1~PeV, and 1~EeV. It follows that the total number of particles scales approximately linear with energy and is given 0.1 times the value obtained using the Heitler approximation~\eqref{Heit} }
\figlab{figA1}
\end{figure}
In~\figref{figA1}, the shower development is plotted at TeV, PeV and EeV energies. Furthermore, we plot the maximum number of particles as given by the Heitler approximation obtained in~\eqref{Heit}. It follows that using the Heitler approximation, the number of particles at the shower maximum is overestimated by a factor of 10. The shower length shown in~\figref{fig1}, defined as the length it takes for the shower to reach its maximum corresponds well to length obtained by the NKG parameterization. 

\section{The radial particle distribution}
For the radial particle distribution we will modify the parameterization obtained in~\cite{Wer12} for cosmic-ray-induced air showers. This parameterization is given as function of the radial distance $r$ in meters. Assuming radial symmetry $w(r)=2\pi r\; w(\vec{r})$, the radial particle distribution is parameterized as,
\beq
w(r)=\frac{\Gamma(4.5-s)}{\Gamma(s)\Gamma(4.5-2s)}\left(\frac{r}{r_0}\right)^{s-1}\left(\frac{r}{r_0}+1\right)^{s-4.5},
\eeq
where $s=1$ corresponds to the shower age at the shower maximum. The second parameter is the distance parameter $r_0$, which for air is approximately given by $r_0^{air}\approx80$~m. From~\eqref{age} it follows that the shower age parameter $s$ is a function of depth $X$, and $s=1$ at the shower maximum in both ice and air. The distance parameter $r_0^{air}$ is given in meters. To convert this distance to ice, we first have to convert it to radiation length,
\bea
X^{air}&=&r_0^{air}\cdot\rho^{air}\nonumber\\
&=& 6.4\; \mathrm{g/cm^2}.
\eea
Where we approximated the air density $\rho^{air}=8\cdot10^{-4}\;\mathrm{g/cm^3}$ at 4~km height corresponding to the position of the air shower maximum. Since the electron radiation length in air $X_0^{air}=36.7\;\mathrm{g/cm^2}$ differs only slightly from its value in ice $X_0^{ice}=36.08\;\mathrm{g/cm^3}$, we assume $X^{air}=X^{ice}$. We can now easily convert $X^{ice}$ into the parameter $r_0^{ice}$. By assuming an ice density of $\rho_{ice}=0.92\;\mathrm{g/cm^3}$, we obtain,
\bea
r_0^{ice}&=&X^{ice}/\rho_{ice}\nonumber\\
&=&7.0\;\mathrm{cm}.
\eea

\end{appendix}

\end{document}